\documentclass{elsart} \usepackage{natbib}

\usepackage{epsfig}

\def\bc{\begin{center}} \def\ec{\end{center}} \def\emi{\end{minipage}}
\def\epi{\end{picture}} \def\bearray{\begin{eqnarray}}
\def\eearray{\end{eqnarray}}

\newcommand{\im}{\mathrm{Im\,}} 
\newcommand{\mh}{m_h} \newcommand{\be}{\begin{equation}}
\newcommand{\ee}{\end{equation}} 
 \newcommand\jps{{J/\psi}}
  \newcommand\eps{\epsilon}
\newcommand\epso{\epsilon_{0}}

\usepackage{epsf} \usepackage{graphicx} \usepackage{amsmath}
\usepackage{amssymb}

\newcommand{\nc}{{N_c}} \newcommand{\mq}{{m_{Q}}}
\newcommand{\dn}{{d_{2k}}} \newcommand{\an}{{A_{2k}}}
\def\cO#1{{{\mathcal{O}}}\left(#1\right)}

\newcommand{\beq}{\begin{equation}} \newcommand{\eeq}{\end{equation}}
\newcommand{\beqa}{\begin{eqnarray}} \newcommand{\eeqa}{\end{eqnarray}}

\newcounter{hran} \renewcommand{\thehran}{\arabic{hran}}

\def\bminiG{\setcounter{hran}{\value{equation}}
\refstepcounter{hran}\setcounter{equation}{0}
\renewcommand{\theequation}{\thehran\alph{equation}}\begin{eqnarray}}

\def\emini{\end{eqnarray}\relax\setcounter{equation}{\value{hran}}\renewcommand{\theequation}{
\arabic{equation}}}

\renewcommand{\thefootnote}{\fnsymbol{footnote}}

\begin{document}

\vspace{-2.cm}
\begin{flushright}
October 2004\\  \texttt{hep-ph/0410295}\\ LPTHE-04-026
\end{flushright}

\begin{frontmatter}
\title{Heavy-quarkonium interaction in QCD \\ at finite temperature}

\author[liege,lpthe]{Fran\c{c}ois Arleo\thanksref{email}},
\author[liege]{Joseph Cugnon}, \author[liege]{Yuri Kalinovsky}

\thanks[email]{Corresponding author.\\{\it Email address:}
\texttt{arleo@lpthe.jussieu.fr} (Fran\c{c}ois Arleo).}

\address[liege]{Universit\'e de Li\`ege, Institut de Physique B5 \\
Sart Tilman, 4000 Li\`ege 1, Belgium} \address[lpthe]{LPTHE,
Universit\'e Paris VI \& Paris VII and CNRS,\\  4, Place Jussieu,
75252 Paris cedex 05, France}

\begin{abstract}
We explore the temperature dependence of the heavy-quarkonium
interaction based on the Bhanot~-~Peskin leading order perturbative
QCD analysis. The Wilson coefficients are computed solving the
Schr\"odinger equation in a screened Coulomb heavy-quark
potential. The inverse Mellin transform of the Wilson coefficients
then allows for the computation of the 1S and 2S heavy-quarkonium
gluon and pion total cross section at finite screening/temperature. As
a phenomenological illustration, the temperature dependence of the 1S
charmonium thermal width is determined and compared to recent lattice
QCD results.
\end{abstract}
\end{frontmatter}

\section{Introduction}

The Debye screening between two opposite color charges is clearly seen
in the QCD static potential computed at finite temperature $T$ on the
lattice~\cite{Karsch:1988rj}. Consequently, heavy quark bound states
(which we call $\Phi$) may no longer exist well above the
deconfinement critical temperature $T_c$, of order
200$-$300~MeV~\cite{Karsch:2001vs}. This has made the heavy-quarkonium
suppression in high energy heavy-ion collisions (as compared to
proton-proton scattering) one of the most popular signatures for
quark-gluon plasma formation~\cite{Matsui:1986dk,Bedjidian:2003gd}. On
the experimental side, a lot of excitement came out a few years ago
after the NA50 collaboration reported a so-called ``anomalous''
suppression in the $J/\psi$ channel in the most central lead-lead
collisions ($\sqrt{s} \simeq 17$~GeV) at the CERN
SPS~\cite{Abreu:2000ni}. At RHIC energy ($\sqrt{s} = 200$~GeV), $\jps$
production has been  measured recently by the PHENIX collaboration
although the presently too large statistical and systematic error bars
prevent one from concluding anything yet quantitative from these
data~\cite{Adler:2003rc}.

The NA50 measurements triggered an intense theoretical activity and
subsequently a longstanding debate on the origin of the observed
$\jps$ suppression. However, it became unfortunately rapidly clear
that no definite conclusion could be drawn as long as theoretical
uncertainties exceed by far that of the high statistics data. Indeed,
both the realistic description of the space-time evolution of the hot
and dense medium as well as the interaction of heavy-quarkonia with
the relevant degrees of freedom (let them be pions or gluons) are
required to be known. While the former can be constrained by global
observables, the latter needs to be computed theoretically. Several
approaches have been suggested to determine heavy-quarkonium total
cross sections, from meson exchange~\cite{Matinyan:1998cb} or
constituent quark models~\cite{Martins:1995hd} to the perturbative
framework developed by Bhanot and
Peskin~\cite{Peskin:1979va,Bhanot:1979vb} upon which the present paper
relies. Let us remark in particular that many recent phenomenological
applications have used the latter perturbative $\Phi$~--~gluon cross
section to estimate the heavy-quarkonium dissociation or formation in
heavy-ion collisions~\cite{Xu:1996eb}.

However, although derived from first principles in QCD perturbation
theory, the Bhanot~-~Peskin result describes the interaction of
Coulombic bound states, that is for which the heavy-quark potential is
well approximated by the perturbative one-gluon exchange potential. As
indicated from spectroscopic studies~\cite{Kwong:1987mj}, this may be
too crude an assumption to describe bound states in the charm or
(even) the bottom sector. Furthermore, it does not take into account
the possible effects of the medium on the heavy-quarkonium
interaction. It is the aim of this Letter to explore how the $\Phi$
interaction with gluons and pions gets modified at finite
temperature. The paper is organized as follows. The general framework
is first briefly recalled in Section~\ref{sec:framework}. Our results
are then detailed in Section~\ref{sec:results} while
Section~\ref{sec:discussion} is devoted to a concluding discussion.

\section{Heavy-quarkonium interaction in QCD}\label{sec:framework}

\subsection{Resummation of the leading-twist forward scattering amplitude}

At leading-twist, the forward heavy-quarkonium ($\Phi$) - hadron ($h$)
scattering amplitude $\mathcal{M}_{\Phi\,h}$ is an operator product
expansion of perturbative Wilson coefficients  $\dn$ evaluated in the
heavy-quarkonium state and computable in perturbation theory times
non-perturbative matrix elements in the hadron state. It
reads~\cite{Peskin:1979va}
\begin{equation}\label{eq:lt_amplitude}
\mathcal{M}_{\Phi\,h}(\lambda)=\left( \frac{g^2\, \nc}{16\,\pi}
\right)\, a_0^2\, \sum_{k\ge 1}\,d_{2k}\,\epsilon^{1-2k}\, \langle
h|\,\frac{1}{2}
F^{0\nu}\,(iD^0)^{2k-2}\,F^{\phantom{\nu}0}_\nu\,|h\rangle
\end{equation}
where $a_0$ and $\epsilon$ stand, respectively, for the Bohr radius
and the binding energy for the $\Phi$ system, $g$ the QCD coupling and
$N_c$ the number of colors. Each of the matrix elements $\langle h |
\dots | h \rangle$ in Eq.~(\ref{eq:lt_amplitude}) is proportional to a
traceless fully symmetric rank $2k$ tensor in the spin-averaged hadron
state~\cite{Peskin:1979va}
\begin{equation*}
\Pi^{\mu_1\cdots\mu_{2k}}(p)= p^{\mu_1}\,\dots \,p^{\mu_{2k}} -
\mathrm{trace\,\, terms}
\end{equation*}
where $p^\mu$ is the hadron momentum. The trace terms correspond to
target mass corrections $\cO{\mh^2/\eps^2}$ which are neglected here
as we shall deal only with pions in the present approach. Note that
such corrections were systematically included in
Refs.~\cite{Kharzeev:1996tw,Arleo:2001mp} and proved relevant only
slightly above the threshold for the quarkonium-hadron interaction
process. The matrix elements can be written as
\begin{equation}\label{eq:gluon_matrix_element}
\langle h|\,\frac{1}{2}
F^{0\nu}(iD^0)^{2k-2}F^{\phantom{\nu}0}_\nu|h\rangle =
A_{2k}\,\Pi^{0\cdots 0}(p) = A_{2k}\,\lambda^{2k}
\end{equation}
where $\lambda \equiv p^0$ is the hadron energy in the $\Phi$ rest
frame and the $\an$ coefficients are the Mellin transform of the
unpolarized gluon density $G^h$ in the hadron
target~\cite{Bhanot:1979vb,Arleo:2001mp}
\begin{equation*}
\an = \int_0^1 \frac{dx}{x} \,x^{2k}\, G^h(x).
\end{equation*}
Plugging (\ref{eq:gluon_matrix_element}) in (\ref{eq:lt_amplitude}),
the leading-twist forward scattering amplitude can be written as
\begin{equation}\label{eq:power_series}
\mathcal{M}_{\Phi\,h}(\lambda)= \left( \frac{g^2\, \nc}{16\,\pi}
\right) \,a_0^2\,\epsilon\, \sum_{k\ge
1}\,\dn\,\an\,(\lambda/\epsilon)^{2k}.
\end{equation}
Expressing the Wilson coefficients in terms of their Mellin moments,
\begin{equation*}
\dn = \int_0^1\,\frac{dx}{x} \,\,x^{2 k}\,\tilde{d}(x),
\end{equation*}
the power series (\ref{eq:power_series}) can be conveniently resumed
and continued analytically throughout the whole complex plane of
energies~\cite{Arleo:2001mp}. This allows for the computation of the
imaginary part of the forward scattering amplitude along the real
axis, $\lambda>\eps$,
\begin{equation}\label{eq:im_amplitude}
\im\,\mathcal{M}(\lambda)=\left( \frac{g^2\, \nc}{32}
\right)\,a_0^2\,\epsilon\,\int_{\epsilon/\lambda}^1 \,\frac{dx}{x}\;
G(x)\;\tilde{d}\left(\frac{\epsilon}{\lambda x}\right).
\end{equation}
Dividing Eq.~(\ref{eq:im_amplitude}) by the flux factor $\lambda$
leads to the total heavy-quarkonium cross section via the optical
theorem
\begin{equation}\label{eq:cross_section}
\sigma_{\Phi\,h}(\lambda) =
\frac{1}{\lambda}\,\im\,\mathcal{M}(\lambda) = \int_0^1
dx\,G(x)\,\sigma_{\Phi\,g}(x\lambda),
\end{equation}
where the heavy-quarkonium gluon cross section is defined as
\begin{equation}\label{eq:cross_section_part}
\sigma_{\Phi\,g}(\omega)= \left( \frac{g^2\, \nc}{32}
\right)\,a_0^2\;\frac{\epsilon}{\omega}\;\tilde{d}\left(\frac{\epsilon}{\omega}\right)
\end{equation}
with the gluon energy $\omega = \lambda\,x$ in the $\Phi$ rest frame.

The Wilson coefficients need first to be computed in an arbitrary
heavy-quark potential and later be inverse Mellin transformed in order
to determine the heavy-quarkonium gluon
Eq.~(\ref{eq:cross_section_part}) and hence the heavy-quarkonium
hadron Eq.~(\ref{eq:cross_section}) total cross sections. This task is
carried out in the next Section.

\subsection{Wilson coefficients and inverse Mellin transform}

Resuming all diagrams contributing to leading order in $g^2$ to the
$\Phi$~--~$h$ interaction, Peskin made explicit the heavy-quarkonium
Wilson coefficients~\cite{Peskin:1979va}. They are given
by\footnote{Note that the coefficients~(\ref{eq:wilson}) are a factor
$(\epsilon/\epso)^{2k-1}$ smaller than in
Ref.~\cite{Peskin:1979va}. This difference is because the energy
$\lambda$ is normalized to the binding energy $\eps$ in the amplitude
(\ref{eq:power_series}) and not to the Rydberg energy $\epso$ as
in~\cite{Peskin:1979va}.}
\begin{eqnarray}\label{eq:wilson}
d_{2k} &=& \frac{16\, \pi}{\nc^2 \,a_0^2}\,\epsilon^{2k-1}\,
\langle\phi |\, r^i\,\frac{1}{\left(H_a +
\epsilon\right)^{2k-1}}\,r^j\,|\phi\rangle\nonumber\\ &=&  \frac{16
\pi}{\nc^2} \int \,\frac{d^3
k}{{(2\pi)}^3}\,\frac{1}{3}\,\vert\frac{{\bf
r}}{a_0}\,\psi|^2(k)\,\epsilon^{2k-1} \langle k |\, \frac{1}{{\left(
H_a +\epsilon \right) }^{2k-1}} \,| k \rangle
\end{eqnarray}
where $|\phi\rangle$ and $k$ are respectively the $Q\bar{Q}$ internal
wavefunction and momenta, while $H_s$ ($H_a$) is the internal
Hamiltonian describing the heavy-quarkonium state in a color-singlet
(color-adjoint) state,
\begin{equation*}
H_{s, a} = \frac{k^2}{\mq} + V_{s, a}(r),
\end{equation*}
$\mq$ being the heavy-quark mass and $V_{s, a}$ the heavy-quark
potential. The heavy-quarkonium wave function $\psi(r)$ in coordinate
space and the binding energy $\epsilon$ appearing in
Eq.~(\ref{eq:wilson}) are determined solving the Schr\"odinger equation
\begin{equation}\label{eq:schrodinger}
H_s \,|\phi\rangle \,=\, -\,\epsilon\,\,|\phi\rangle
\end{equation}
in the color singlet potential.

\subsubsection{Coulomb potential}

The leading-twist amplitude (\ref{eq:lt_amplitude}) was determined
assuming the $Q\bar{Q}$ binding potential is well approximated by the
one-gluon exchange Coulomb potential
\begin{eqnarray}\label{eq:coulomb}
\begin{split}
V_{s} &= - \, \frac{g^2\,N_c}{8\,\pi \,r} + \cO{\nc^{-1}},\\ V_{a} &=
\cO{\nc^{-1}},
\end{split}
\end{eqnarray}
in SU($\nc$) gauge theory. To leading order in $\cO{\nc^{-1}}$, $H_a$
is given by the free-particle Hamiltonian and the Wilson
coefficients~(\ref{eq:wilson}) read
\begin{equation*}
d_{2k} =  \frac{16 \pi}{\nc^2} \int \,\frac{d^3
k}{{(2\pi)}^3}\,\frac{1}{3}\,\vert\frac{{\bf
r}}{a_0}\,\psi|^2(k)\,\frac{(\epsilon/\epso)^{2k-1}}{{\left[ (k a_0)^2
+ \epsilon/\epso\right] }^{2k-1}},
\end{equation*}
where we have introduced the Rydberg energy $\epso$ for the $Q\bar{Q}$
system
\begin{equation*}
\epso = \left(\frac{g^2\,\nc}{16\,\pi}\right)^2\,\mq =
\frac{1}{\mq\,a_0^2}
\end{equation*}
Solving the Schr\"odinger equation~(\ref{eq:schrodinger}) gives the
well-known 1S and 2S Coulomb wave functions with the corresponding
binding energies,
\begin{eqnarray*}\label{eq:psi}
\begin{split}
a_0^{3/2} \psi^{(1S)}(r) &= \frac{1}{\sqrt{\pi}}\,\exp \left(-
\frac{r}{a_0}\right)
\quad\qquad\qquad\qquad\quad;\qquad\epsilon_{_{\mathrm{1S}}} = \epso\\
a_0^{3/2}\psi^{(2S)}(r) &= \frac{1}{\sqrt{8 \pi}}\,\left(1-\frac{r}{2
a_0}\right)\,\exp \left(-\frac{r}{2 a_0}\right)\qquad
;\qquad\epsilon_{_{\mathrm{2S}}} = \epso / 4
\end{split}
\end{eqnarray*}
which eventually allows for the computation of the Wilson
coefficients~\cite{Peskin:1979va}
\begin{eqnarray}\label{eq:dn}
\begin{split}
d_n^{(1S)}&=\int_0^1 \frac{dx}{x} \,x^n\,\frac{16^3}{3 N_c^2}
x^{5/2}(1-x)^{3/2},\\ d_n^{(2S)}&=\int_0^1 \frac{dx}{x}
\,x^n\,\frac{16 \times 16^3}{3 N_c^2} x^{5/2}(1-x)^{3/2}(1-3x)^2.
\end{split}
\end{eqnarray}
From Eqs.~(\ref{eq:cross_section_part}) and (\ref{eq:dn}), the
expression for the inverse Mellin transform $\tilde{d}(x)$ is
straightforward and one gets directly~\cite{Bhanot:1979vb}
\begin{equation}\label{eq:coulomb1s}
\sigma_{\Phi^{(1S)}\,g}(\omega)=\frac{16^2 g^2}{6N_c}\,a_0^2\,
\frac{(\omega/\epsilon_{_{\mathrm{1S}}}-1)^{3/2}}{(\omega/\epsilon_{_{\mathrm{1S}}})^5}\,
\theta(\omega-\epsilon_{_{\mathrm{1S}}}),
\end{equation}
for 1S states and~\cite{Arleo:2001mp}
\begin{equation}\label{eq:coulomb2s}
\sigma_{\Phi^{(2S)}\,g}(\omega)=16\times\frac{16^2 g^2}{6N_c}\,a_0^2\,
\frac{(\omega/\epsilon_{_{\mathrm{2S}}}-1)^{3/2}(\omega/\epsilon_{_{\mathrm{2S}}}-3)^2}{(\omega/\epsilon_{_{\mathrm{2S}}})^7}\,\theta(\omega-\epsilon_{_{\mathrm{2S}}})
\end{equation}
for 2S states. Note that these expressions were also obtained by 
Kim, Lee, Oh and Song from the QCD factorization property combined with the
Bethe-Salpeter amplitude for the heavy-quark bound state, which
allowed them to include relativistic and next-to-leading order
corrections~\cite{Oh:2001rm}.

\subsubsection{Screened Coulomb potential}

\begin{figure}
\begin{center}
\includegraphics[width=12.0cm]{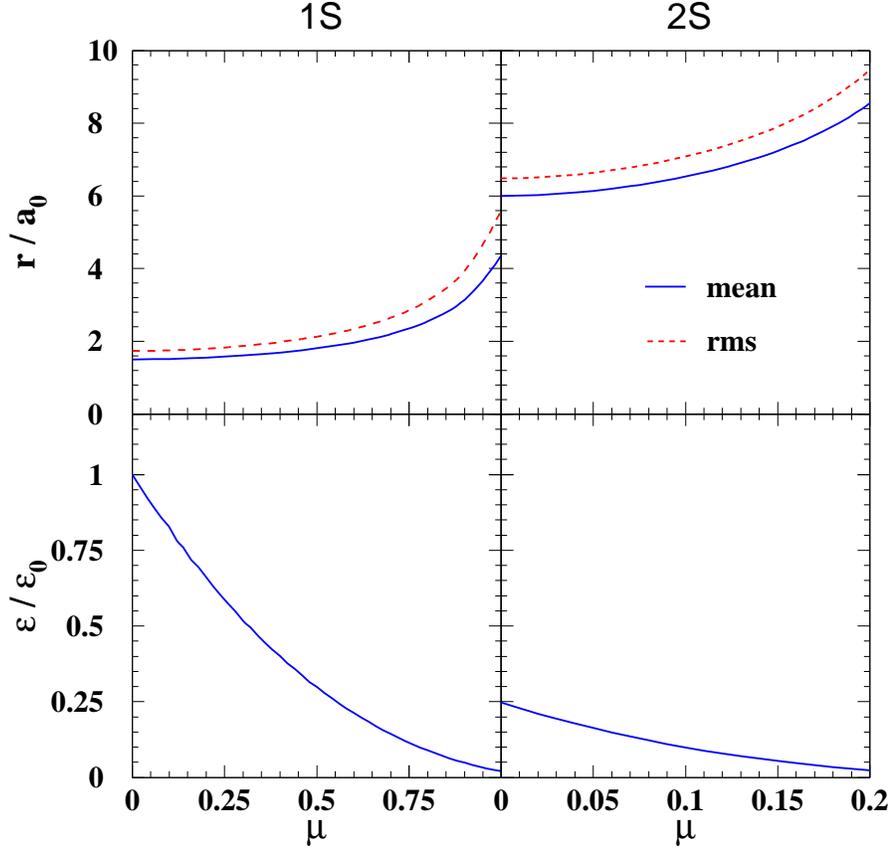}
\end{center}
    \caption{{\it Top:} Mean radius ({\it solid}) and root mean square
    radius ({\it dashed}) of the 1S ({\it left}) and 2S ({\it right})
    heavy-quarkonium states as a function of $\mu$. {\it Bottom:} 1S
    ({\it left}) and 2S ({\it right}) binding energy as a function of
    $\mu$.}
\label{fig:radii}
\end{figure}

As stressed in the Introduction, the above formulas may serve as an
important input to estimate the heavy-quarkonium dissociation process
$\Phi + g \to Q + \bar{Q}$ (or the detailed balance process) in a hot
gluon or pion gas formed in high energy heavy-ion collisions. We would
like here to go one step further and to discuss possible medium
modifications to these total cross sections. Medium effects will be
modeled at the level of the heavy-quarkonium potential by considering
a screened Coulomb potential (Yukawa type) characterized by a
dimensionless screening parameter $\mu$,
\begin{eqnarray}\label{eq:screenedcoulomb}
\begin{split}
V_{s} &= - \, \frac{g^2\,N_c}{8\,\pi \,r}\, \exp \left({- \mu\,r /
a_0} \right),\\ V_{a} &= 0.
\end{split}
\end{eqnarray}
Solving the Schr\"odinger equation (\ref{eq:schrodinger}) using the
potential~(\ref{eq:screenedcoulomb}), the wave functions and binding
energies for 1S and 2S states are determined and the corresponding
Wilson coefficients~(\ref{eq:wilson})  are computed numerically
subsequently. For the illustration, we plot in Figure~\ref{fig:radii}
the typical size (mean and root mean square radii, {\it top}) as well
as the binding energy ({\it bottom}) for the 1S ({\it left}) and 2S
({\it right}) $\Phi$ states.
Finally, the inverse Mellin transform
\begin{equation*}
\tilde{d}(x) = \frac{1}{2\,i\,\pi}\,\int_{c - i \infty}^{c + i \infty}
\,dz \,\,x^{-z}\,d(z)
\end{equation*}
$c$ being a real constant, is performed thus giving access to the
medium-modified total cross sections.

\section{Results}\label{sec:results}

\subsection{Finite screening}

Before discussing the results, both the Bohr radius $a_0$ and the
Rydberg energy $\epso$ in the charmonium and bottomonium channel need
to be fixed. Assuming both the 1S and 2S states to be Coulombic, the
heavy quark mass $m_Q$ and the Rydberg energy $\epsilon_0$ can be
determined from the 1S and 2S heavy-quarkonium masses. One then
obtains~\cite{Arleo:2001mp}
\begin{eqnarray*}
\epsilon_{0 c} = 0.78\;\mathrm{GeV}&\;\;,\;\;&a_{0 c}^{-1} = 1.23
\;\mathrm{GeV}, \\ \epsilon_{0 b} = 0.75\;\mathrm{GeV}&\;\;,\;\;&a_{0
b}^{-1} = 1.96 \;\mathrm{GeV}.
\end{eqnarray*}

\begin{figure}[h]
\begin{center}
\includegraphics[width=15.0cm]{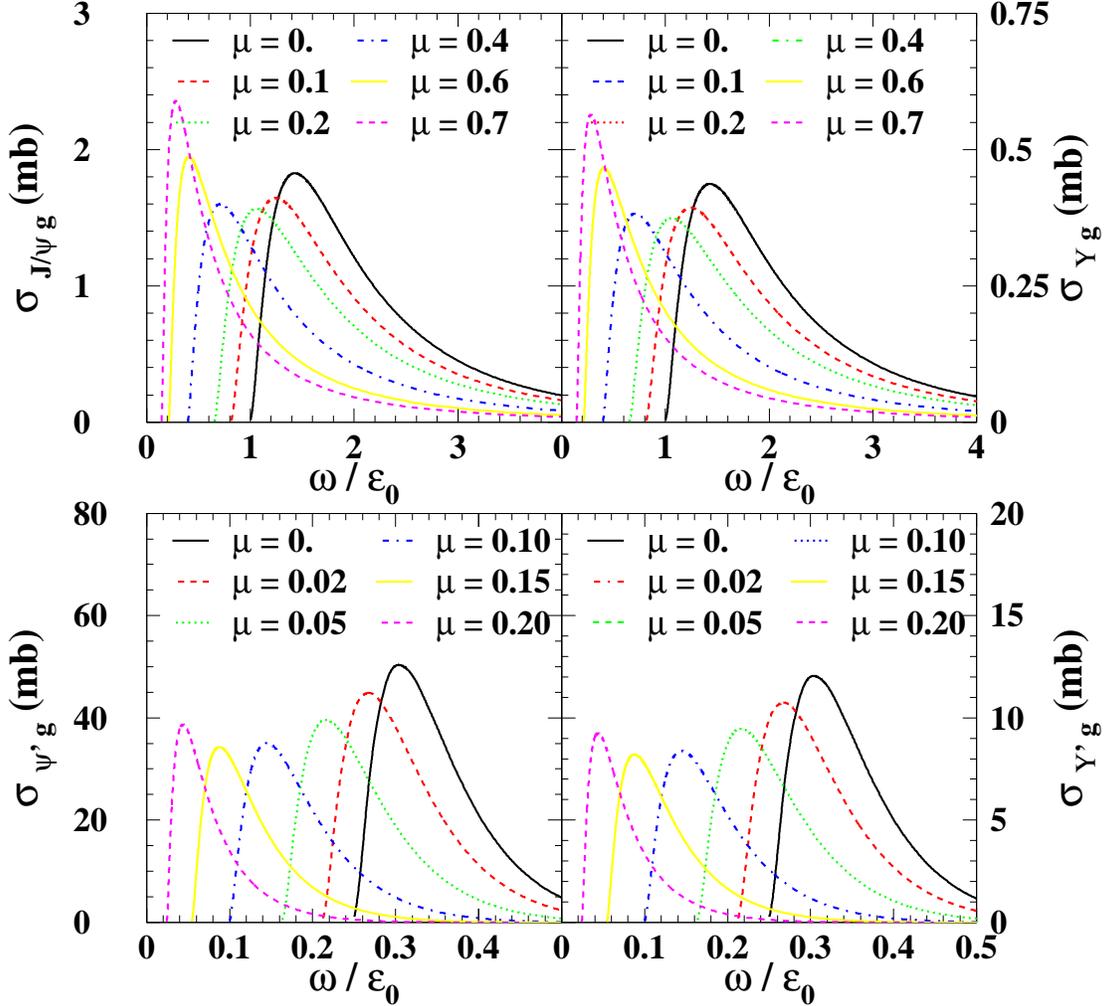}
\end{center}
    \caption{1S ({\it top}) and 2S ({\it bottom}) charmonium ({\it
    left}) and bottomonium ({\it right}) gluon total cross section as
    a function of the gluon energy $\omega$ for various values of the
    screening parameter $\mu$.}
\label{fig:cross_sect_part}
\end{figure}

Using the above (not too hard) scales, the 1S ({\it top}) and 2S ({\it
bottom}) heavy-quarkonium gluon dissociation cross sections are
computed in Figure~\ref{fig:cross_sect_part} as a function of the
gluon energy $\omega$ for various values of the screening parameter
$\mu$. The dominant effect of the screened heavy quark potential is
the decrease of the 1S (respectively, 2S) heavy-quarkonium binding
energy from $\epsilon_0$ (respectively, $\epsilon_0/4$) to $\epsilon$
which leads to a lower threshold for the inelastic process. The medium
modifications of the $\Phi$~--~gluon total cross sections are
nevertheless not only due to the smaller binding energy, yet the
characteristic shapes of the cross sections are reminiscent to what is
already known for pure Coulombic states, $\mu = 0$
(Figure~\ref{fig:cross_sect_part}, {\it solid}). We checked for
instance that the Wilson coefficients get somehow modified at finite
screening and consequently the partonic cross sections do not simply
scale as $\omega/\epsilon$ in~Eq.~(\ref{eq:coulomb1s}). This is a
strong indication that cross sections cannot be deduced with a simple
rescaling of the binding energy from $\epso$ to $\epsilon$  to mimic
medium effects in the heavy-quarkonium dissociation process. Finally,
the significant increase of the 1S partonic cross sections at large
screening is particularly noticeable as the dipole size gets
larger. However, as discussed later, reliable calculations require the
space-time scales to remain small which prevent one from taking
arbitrarily large screening parameter values, at least when
considering such ``light'' heavy quarks\footnote{According
to~\cite{Bhanot:1979vb}, the assumption of heavy-quark Coulombic bound
states should be appropriate for more than 25~GeV heavy quark mass.}.
Moreover, since the heavy-quark potential in the original QCD analysis
needs to be Coulomb-like, the screening parameter $\mu$ in the model
Eq.~(\ref{eq:screenedcoulomb}) should remain small as compared to one.

\begin{figure}[h]
\begin{center}
\includegraphics[width=15.0cm]{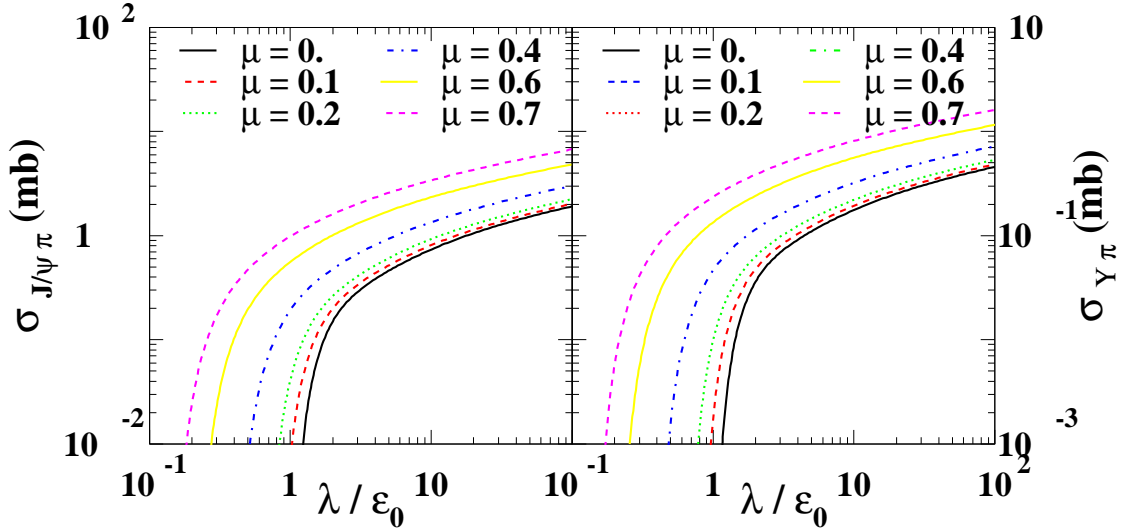}
\end{center}
    \caption{$J/\psi-\pi$ ({\it left}) and $\Upsilon-\pi$ ({\it
    right}) total cross section as a function of the pion energy
    $\lambda$ for various values of the screening parameter $\mu$.}
\label{fig:cross_sect_had}
\end{figure}

Let us now discuss the heavy-quarkonium hadron cross section. Since
heavy-quarkonia plunged into the hot medium are most likely to
interact with pions, we shall only consider the $\Phi$~--~$\pi$
channel and choose the GRV LO parameterization for the gluon
distribution\footnote{These should be evaluated at a factorization
scale $\epsilon$. We take in the following a frozen scale
$\epsilon_0$.} in the pion~\cite{Gluck:1991ey}. The $\jps$~--~$\pi$
and $\Upsilon$~--~$\pi$ cross sections are computed in
Figure~\ref{fig:cross_sect_had} as a function of the pion energy
$\lambda$. Again, the threshold for the process, located at
$\lambda=\epsilon$, gets shifted to lower values leading to a strong
modification of the heavy-quarkonium pion interaction in this
region. At high energy, small $x=\cO{\epsilon/\lambda}$ gluons
dissociate heavy-quarkonia, thereby increasing the $\Phi$~--~$\pi$
cross section by a factor $\left(\epso/\eps\right)^\delta$ where
$\delta\simeq 0.3$ governs the rise of the gluon distribution at small
$x$, $x G(x) \propto x^{-\delta}$~\cite{Aid:1996au}.

\subsection{Finite temperature}

The $\Phi$ interaction with gluons and pions has been computed so far
using a heavy-quark screened Coulomb potential characterized by one
parameter $\mu$. Interpreting $\mu$ as the screening mass in a gluon
plasma, the model for the finite temperature $Q\bar{Q}$ potential now
looks like
\begin{equation}\label{eq:screenedcoulomb_run}
V_{s} = - \, \frac{g^2(r, T) \,N_c}{8\,\pi \,r}\, \exp \left({-
m_D(T)\,r}\right)
\end{equation}
At short distance and/or low temperature, we shall consider a frozen
coupling constant
\begin{equation}\label{eq:running_short}
g^2(r, T) = g^2 \qquad {\mathrm{for}}\ r\,T\ll \Lambda
\end{equation}
and recover the Coulomb potential behavior~(\ref{eq:coulomb}), while
the QCD coupling starts to run with $T$ at large distance and/or high
temperature. At two loops, we have
\begin{equation}\label{eq:running_long}
g^2(r, T) \equiv \tilde{g}^2(T) =
\Biggr(\frac{11}{8\pi^2}\ln\left(\frac{2 \pi
T}{\Lambda_{\overline{\mathrm{MS}}}}\right) +
\frac{51}{88\pi^2}\ln\left[2\ln\left(\frac{2 \pi
T}{\Lambda_{\overline{\mathrm{MS}}}}\right)\right]\Biggr)^{-1}\qquad
{\mathrm{for}}\ r\,T \gg \Lambda
\end{equation}
with $T_c / \Lambda_{\overline{\mathrm{MS}}} =
1.14$~\cite{Kaczmarek:2004gv}. The Debye mass $m_D$ is related to the
temperature through the leading-order perturbative result,
\begin{equation*}
m_D(T) = \tilde{g}(T)\,T.
\end{equation*}
The $\Lambda$ dimensionless parameter introduced in
Ref.~\cite{Kaczmarek:2004gv} separates somewhat arbitrarily the short
from the long distance physics at finite temperature. Fitting pure
gauge SU(3) heavy quark potential, they obtained the empirical value
$\Lambda = 0.48 \,\mathrm{fm} \times
T_c$. Following~\cite{Kaczmarek:2004gv}, we shall take the 2-loop
running coupling~(\ref{eq:running_long}) rescaled by 2.095 and
interpolate smoothly between the short and long distance
regime\footnote{Similar results are obtained using the one loop
running coupling with an appropriate rescaling.}.
\begin{figure}[h]
\begin{center}
\includegraphics[width=15.0cm]{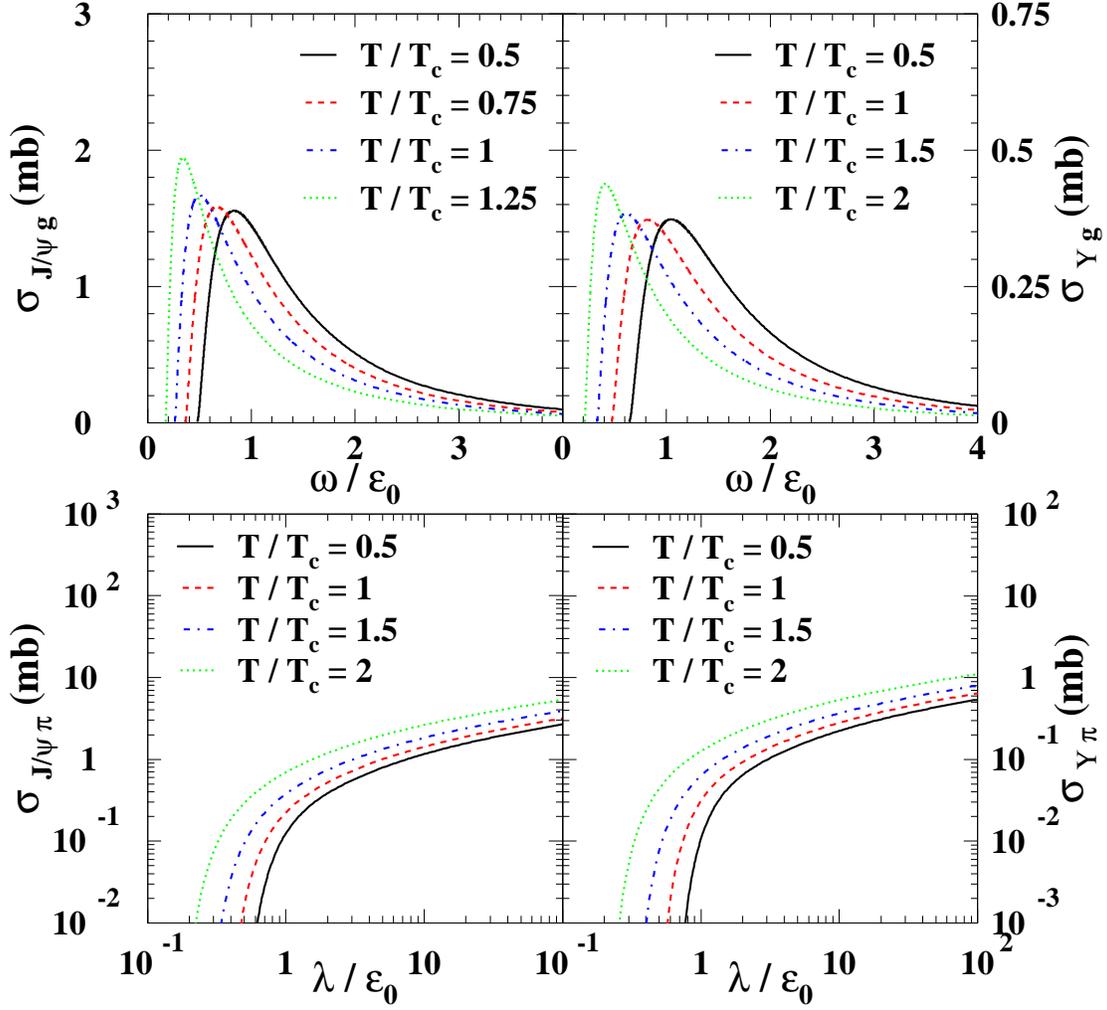}
\end{center}
    \caption{$J/\psi$ ({\it left}) and $\Upsilon$ ({\it right}) total
    cross sections with gluons ({\it top}) and pions ({\it bottom}) at
    various temperatures.}
\label{fig:cross_sect_temp}
\end{figure}

The partonic and hadronic $\jps$ and $\Upsilon$ cross sections are
computed in Figure~\ref{fig:cross_sect_temp} for several temperatures
in units of the critical temperature for deconfinement, $T_c =
270$~MeV in SU(3) pure gauge theory~\cite{Karsch:2001vs}. The
temperatures selected for the bottomonium system are chosen to be
slightly higher than those for the charmonium system since the larger
bottom quark mass (hence, smaller size) probes more efficiently hotter
QCD media~\cite{Digal:2001ue}.

The effects of the running coupling in
Eq.~(\ref{eq:screenedcoulomb_run}) being quite small, rather similar
features at finite temperature and at finite screening are
observed. In particular, the charmonium binding energy (hence the
inelastic threshold) drops by a factor of two already at $T/T_c=0.5$
and thus affects dramatically the $\jps$ interaction in the vicinity
of the threshold. At higher temperature, the $\jps$~--~gluon cross
section is significantly enhanced at small gluon energy due to the
larger charmonium size. The $\jps$~--~$\pi$ cross section is also
somewhat modified with a magnitude increasing noticeably with the
temperature. Moving to the bottom sector
(Figure~\ref{fig:cross_sect_temp}, {\it right}), the $\Upsilon$ cross
sections exhibit the same general characteristics yet the medium
effects at a given temperature prove much less pronounced from the
smaller bottomonium size.

At high temperature, heavy-quarkonium interaction can not be described
by short-distance techniques (see Fig.~\ref{fig:radii}) and our
predictions are not valid any longer. On top of that, the process
described here is the heavy-quarkonium dissociation by {\it hard}
gluons as opposed to the soft gluons which only affect its
properties. Therefore, our calculations should be valid as long as the
Debye mass is kept smaller than the heavy-quarkonium Rydberg energy,
$m_D(T)~\lesssim~\epso$. This condition is fulfilled provided the bath
temperature is smaller than 350~MeV. Above that scale, the screened
exchanges are able to dissociate the bound states, the factorization
between the heavy-quarkonium physics and the external gluon field is
broken and the above QCD picture loses its significance.

\subsection{$J/\psi$ spectral function width}

The former results indicate that Debye screening effects may play an
important role in the heavy-quarkonium dissociation by incoming gluons
or pions. In order to illustrate how medium modifications could affect
the $\Phi$ suppression in heavy-ion collisions, we compute in this
section the 1S charmonium thermal width $\Gamma_{_{\jps}}$ (or
equivalently its lifetime, $\tau_{_{\jps}} = \Gamma_{_{\jps}}^{-1}$)
in a hot gluon bath. Assuming the $\jps$ suppression is only due to
the gluon dissociation process, the width can be written
\begin{equation*}
\Gamma_{_{\jps}}(T) =
\frac{1}{2\,\pi^2}\,\int_0^{\infty}\,\omega^2\,d\omega\,\sigma_{\jps
g}(\omega, T)\,n_g(\omega, T)
\end{equation*}
where $n_g(\omega, T) = 2 (N_c^2-1)\bigr/ (\exp\left(\omega/T\right) -
1)$ is the gluon density in a gluon gas in thermal equilibrium.
\begin{figure}[h]
\begin{center}
\includegraphics[width=8.0cm]{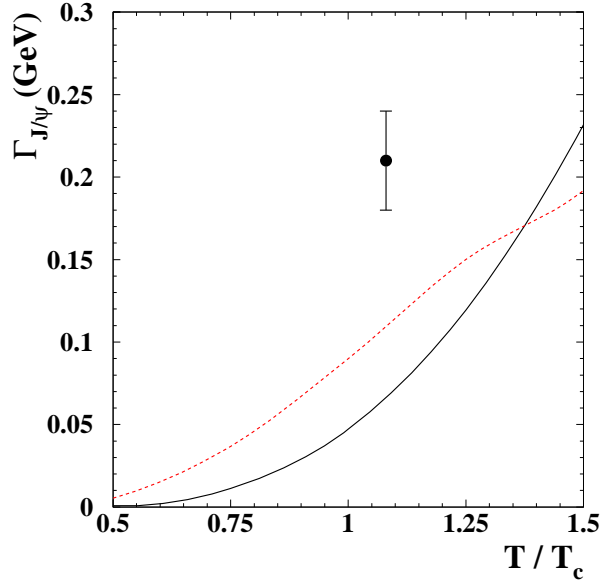}
\end{center}
    \caption{$J/\psi$ thermal width as a function of the temperature
    with ({\it dotted}) and without ({\it solid}) modifications of the
    heavy-quark potential. The lattice data point obtained in
    Ref.~\cite{Umeda:2002vr} is also shown for comparison.}
\label{fig:width}
\end{figure}

The thermal width is computed in Figure~\ref{fig:width} as a function
of the temperature $T$ assuming the vacuum ({\it solid}) and the
in-medium ({\it dashed}) $\jps$~--~gluon cross section. At small
temperature, $T \ll \eps$, most gluons are not sufficiently energetic
to dissociate $\jps$ states and the width remain small as the phase
space selected by the $\jps$ gluon threshold is restricted. When the
medium gets warmer, more and more gluons are able to interact
inelastically with the $\jps$, hence the thermal width
increases. Interestingly enough, the in-medium $\jps$ thermal width
proves larger by a factor of two or more up to $T=T_c$ due to the
lower threshold in the medium modified cross sections. At even higher
temperature, the medium modified result becomes smaller to that in the
vacuum since {\it dissociating} gluons (with $\omega$ of order
$\epsilon$)  grow scarce. Also plotted in Figure~\ref{fig:width} is
the $\jps$ width computed recently on the lattice at finite
temperature in the quenched
approximation~\cite{Umeda:2002vr}. Although a significant discrepancy
remains between our calculations and the lattice data point, it is
interesting to note that adding medium effects tends to reduce the
disagreement, whose origin is not clarified.

\section{Concluding discussion}\label{sec:discussion}

Before summarizing our main results, we would like to discuss the
limitations of our approach. The starting point of the calculation is
the forward scattering amplitude $\mathcal{M}_{\Phi\,h}$ originally
derived for Coulomb bound states. To go beyond this one-gluon exchange
picture would require to include light quark loops in the
diagrammatics, to which the soft gluon source may couple, that we have
not attempted. However, as conjectured in~\cite{Bhanot:1979vb}, it is
appealing to guess that the generic dipole coupling appearing to
leading order in $g^2$ in the heavy-quarkonium Wilson
coefficients~(\ref{eq:wilson}) survives perturbative and
non-perturbative modifications of the $Q\bar{Q}$ binding
potential. Therefore we believe that taking the literal expression for
the Coulomb states Wilson coefficients and compute them in a screened
Coulomb potential appears sensible, at least as long as the screening
remains reasonable, $m_D \, a_0 \ll 1$. This is certainly the case
when the temperature is kept small as compared to the heavy quark
mass. In that sense, the smallness of the charm and bottom quark mass
as compared to the non-perturbative scale of QCD indeed remains a
problematic issue. As we have seen, typical space time scale becomes
increasingly larger with the temperature, thus strongly limiting our
confidence in the high temperature regime. Finally, one should keep a
clear factorization between the gluon source and the heavy-quarkonium
swimming in the gluon bath. We have seen that such a separation should
be achieved as long as the Debye mass is small as compared to the
bound state Rydberg energy, that is for temperatures $T \lesssim
350$~MeV.

We presented a numerical calculation of the heavy-quarkonium cross
section with gluons and pions, taking into account the possible
medium-modifications of the heavy-quark potential at finite
temperature. Such a work can therefore be useful to estimate
heavy-quarkonium production in high energy heavy-ion collisions. In
particular, we feel it would be interesting to explore the
phenomenological consequences of such corrections comparing them to
present calculations based on the vacuum heavy-quarkonium
interaction. Finally, this very framework could be applied to study
the $\Phi$ interaction using a variety of realistic heavy quark
(confining) potentials currently used in charmonium and bottomonium
spectroscopy~\cite{Kwong:1987mj} to describe more accurately, although
further away from the perturbative requirement, heavy-quarkonium
interaction with gluons and hadrons.

\section*{Acknowledgements}

This work was supported the Belgian "Institut Interuniversitaire des
Sciences Nucleaires" foundation.

\bibliographystyle{unsrt}

\begin{thebibliography}{99}

\bibitem{Karsch:1988rj} F.~Karsch, H.W. Wyld, Phys. Lett. {\bf B213},
505 (1988);\\ F.~Karsch, J. Phys. {\bf G30}, 887 (2004).

\bibitem{Karsch:2001vs} F. Karsch, Nucl. Phys. {\bf A698}, 199 (2002).

\bibitem{Matsui:1986dk} T.~Matsui, H.~Satz, Phys. Lett. {\bf B178},
416 (1986).

\bibitem{Bedjidian:2003gd} For a review, see M. Bedjidian {\em et
al.}, CERN-2004-009, \texttt{hep-ph/0311048}.

\bibitem{Abreu:2000ni} NA50 collaboration, M.~C. Abreu {\em et~al.},
Phys. Lett. {\bf B477}, 28 (2000).

\bibitem{Adler:2003rc} PHENIX collaboration, S.~S. Adler {\em et~al.},
Phys. Rev. {\bf C69}, 014901 (2004).

\bibitem{Matinyan:1998cb} S.G. Matinyan, B. M\"uller, Phys. Rev. {\bf
C58}, 2994 (1998);\\ Y. Oh, T. Song, S.H. Lee, Phys. Rev. {\bf C63},
034901 (2001);\\ L.~Maiani, F.~Piccinini, A.D. Polosa, V.~Riquer,
Nucl. Phys. {\bf A741}, 273 (2004).

\bibitem{Martins:1995hd} K.~Martins, D.~Blaschke, E.~Quack,
Phys. Rev. {\bf C51}, 2723 (1995);\\ C.-Y. Wong, E.S. Swanson,
T. Barnes, Phys. Rev. {\bf C62}, 045201 (2000).

\bibitem{Peskin:1979va} M.E.~Peskin, Nucl. Phys. {\bf B156}, 365
  (1979).

\bibitem{Bhanot:1979vb} G.~Bhanot, M.E.~Peskin, Nucl. Phys. {\bf
B156}, 391 (1979).

\bibitem{Xu:1996eb} X.-M.~Xu, D.~Kharzeev, H.~Satz, X.-N.~Wang,
Phys. Rev. {\bf C53}, 3051 (1996);\\ R.L. Thews, M. Schroedter,
J. Rafelski, Phys. Rev. {\bf C63}, 054905 (2001);\\ L.~Grandchamp,
R.~Rapp, Phys. Lett. {\bf B523}, 60 (2001);\\ A. Polleri, T. Renk,
R. Schneider, W. Weise, Phys. Rev. {\bf C70}, 044906 (2004).

\bibitem{Kwong:1987mj} W. Kwong, J. Rosner, C. Quigg,
Ann. Rev. Nucl. Part. Sci. {\bf 37}, 325 (1987).

\bibitem{Kharzeev:1996tw} D.~Kharzeev, H.~Satz, A.~Syamtomov,
G.~Zinovjev, Phys. Lett. {\bf B389}, 595 (1996).

\bibitem{Arleo:2001mp} F.~Arleo, P.-B. Gossiaux, T.~Gousset,
J.~Aichelin, Phys. Rev. {\bf D65}, 014005 (2002).

\bibitem{Oh:2001rm} Y.-S. Oh, S. Kim, S.H. Lee, Phys. Rev. {\bf C65},
067901 (2002);\\
T. Song, S.H. Lee, \texttt {hep-ph/0501252}.

\bibitem{Gluck:1991ey} M.~Gl\"uck, E.~Reya, A.~Vogt, Z. Phys. {\bf
C53}, 651 (1992).

\bibitem{Aid:1996au}  H1 collaboration, S. Aid {\em et al.},
Nucl. Phys. {\bf B470}, 3 (1996).

\bibitem{Kaczmarek:2004gv} O.~Kaczmarek, F.~Karsch, F.~Zantow,
P.~Petreczky, Phys. Rev. {\bf D70}, 074505 (2004).

\bibitem{Digal:2001ue} S.~Digal, P.~Petreczky, H.~Satz,
Phys. Rev. {\bf D64}, 094015 (2001).

\bibitem{Umeda:2002vr} T. Umeda, K. Nomura, H. Matsufuru,
\texttt{hep-lat/0211003}.

\end{thebibliography}

\end{document}